\documentclass[10pt,a4paper,twocolumn]{IEEEtran}
\usepackage{amssymb}
\usepackage[english]{babel}
\usepackage{graphicx}
\usepackage{latexsym}
\usepackage{varioref}
\usepackage{epsfig}

\usepackage{multirow}
\usepackage{array}
\usepackage{hhline}
\usepackage{dcolumn}
\usepackage{tabularx}
\usepackage{algorithm}
\usepackage{algorithmic}
\usepackage{calc}
\usepackage{subfigure}
\usepackage{amssymb,amsmath,epsfig}
\usepackage{slashbox}
\usepackage[usenames]{color}

\usepackage{t1enc}
\usepackage{pstricks,pst-node,psfrag,pst-plot}
\usepackage{graphicx,parskip,amsbsy}
\usepackage{boxedminipage}
\usepackage{indentfirst}
\usepackage{booktabs}
\usepackage{amssymb}
\usepackage{multirow}
\usepackage{savesym}
\usepackage{amsmath}
\savesymbol{iint}
\usepackage{txfonts}
\restoresymbol{TXF}{iint}

\newcommand{\bs}{\boldsymbol}

\newcommand{\bsH}{{\bs H}}

\newcommand{\bsr}{{\bs r}}

\newcommand{\rem}[1]{}
\def\BibTeX{{\rm B\kern-.05em{\sc i\kern-.025em b}\kern-.08em
    T\kern-.1667em\lower.7ex\hbox{E}\kern-.125emX}}

\renewcommand{\baselinestretch}{0.98}

\renewcommand{\baselinestretch}{1}
\begin{document}

\pagestyle{plain}
\title{Extension of ITU IMT-A Channel Models for Elevation Domains and Line-of-Sight Scenarios}
\author{\vspace{0cm}\authorblockN{Zhimeng Zhong$^{1}$ , Xuefeng Yin$^{2}$ , Xin Li$^{1}$ and Xue Li$^{1}$}
\authorblockA{
\\$^{1}$Huawei Technology Company, Xi'an, China
\\$^{2}$School of Electronics and Information Engineering, Tongji University, Shanghai, China
\\Email: zmzhong@huawei.com and yinxuefeng@tongji.edu.cn\vspace*{-4ex}
               }\thanks{This work is supported Huawei Research Project [Channel Characteristics Map for Operational Wireless Communication Netowrks].}}

\maketitle
\thispagestyle{plain} % This is for removing the first page number

%\tableofcontents
\begin{abstract}
In this contribution, the 3-dimensional (3D) channel characteristics, particularly in the elevation domains, are extracted through measurements in typical urban macro and micro environments in Xi'an China. Stochastic channel model parameters are obtained based on the high-resolution multi-path parameter estimates. In addition, a modified spatial channel model (SCM) for the line-of-sight (LoS) scenario is proposed where the LoS polarization matrix is parameterized in accordance with the reality. Measurement results justify the reasonability of the proposed model. These works significantly improve the applicability of the ITU SCM models in realistic 3D channel simulations.
 \end{abstract}
\vspace*{-2ex}

\IEEEpeerreviewmaketitle
\section{Introduction}
Multiple-input Multiple-ouput (MIMO) enhancements are proposed in the 3rd Generation Partnership Project (3GPP) Releases 10 and 11 to support eNodeB (eNB) antenna configurations capable of adaptation in azimuth. Recently researches in enhancing system performance through the use of antenna systems having a two-dimensional array structure has been paid lots of attention in order to provide more spatial degree of freedom in both elevation and azimuth domains \cite{RP-121412}. Additional control in the elevation dimension enables new transmission techniques by using the 3-dimensional(3D)-beamforming or with the massive MIMO, such as the sector-specific elevation beamforming (e.g., adaptive control over the vertical pattern beamwidth and/or downtilt), advanced sectorization in the vertical domain, and user-specific elevation beamforming techniques. In order to evaluate the performance of the systems utilizing these techniques, new channel models characterizing channels in both vertical and horizontal dimensions in various environments with different user locations are necessary.

Existing directional channel models as proposed in 3GPP TR25.996 \cite{TR25996}, the IMT-Advanced standards \cite{M2135} and in the WINNER reports \cite{WINNER} have been widely adopted for development of wireless communication systems. However, with respect to the elevation characteristics of the radio channel, these models are considered to be insufficient. For example, in the WIINNER+ project \cite{WINNER+}, 3D channel modeling methodology and parameters were proposed based on literature survey rather than extensive measurements. Thus, the applicability of the models suggested are questionable due to some obvious errors, such as that the cross-correlation matrix of large scale parameters (LSPs) is non-positive definite. The elevation characteristics, such as mean elevation angular spread (MEAS), extracted based on measurements have also been reported in literature \cite{Kalliola1,Kalliola2,Ericsson1}. However, these models focus on the marginal elevation characteristics, rather than the joint characteristics in both azimuth and elevation, and thus, are lack of applicability for 3D MIMO system-level evaluation.

In this contribution, 3D channel measurement campaigns conducted by Huawei Propagation Research Group are described, which have been performed in different urban scenarios with the objective of extracting the full 3D channel characteristics particularly at the eNB. New measurement results for the statistics of the LSPs, and the cross-correlation matrix thereof are elaborated. Furthermore, a questionable setting specified in the WINNER+ SCM models for the polarization matrix of the line-of-sight (LoS) path is corrected and verified with measurement data.

The rest of the paper is organized as follows. Section \ref{sect:experimentalsettings} introduces the measurement setup and the environments where the measurements were carried out. In Section \ref{sect:resultsanalysis}, the channel characteristics extracted for the urban macro (UMA) and urban micro (UMI) scenarios are described and the updated values for the ITU IMT-A model parameters are reported. In Section \ref{sect:loscoffecients}, a mistake in the SCM, WINNER+, and ITU IMT-Advanced models for the LoS scenario is discussed, and the correction is proposed. Finally conclusive remarks are provided in Section \ref{Sect:conclusions}.

\section{Measurement setup and environments}\label{sect:experimentalsettings}
%\subsection{3D channel sounder and measurement environment}
\begin{figure}\vspace*{2ex}
\centering
\includegraphics[width=7.5cm]{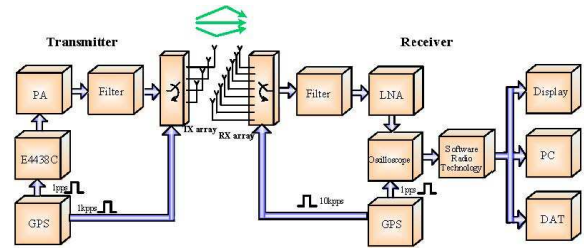}
%\rput[tl](-6cm,3.7cm){\small (a) the Tx}
%\rput[tl](-2.5cm,3.7cm){\small  (b) the Rx}
\caption{The system diagram of the measurement set-up}\label{figure1}
\end{figure}
\begin{figure}
\centering
\includegraphics[width=7.5cm]{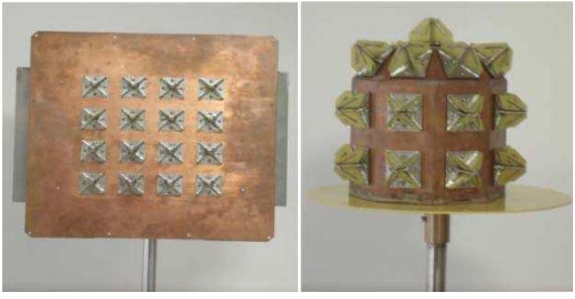}
%\rput[tl](-6cm,3.7cm){\small (a) the Tx}
%\rput[tl](-2.5cm,3.7cm){\small  (b) the Rx}
\caption{Antenna arrays at eNB (left) and UE (right) sides}\label{figure2}
\end{figure}

A diagram for the measurement setup is illustrated in Figure \ref{figure1}, where the Agilent E4438C signal generator is applied as an eNB, generating a pseudo-noise signal with $35$MHz bandwidth and center frequency of $2.6$ GHz. Figure \ref{figure2} depicts the photographs of the antennas used in the eNB and the user equipment (UE). The eNB was equipped with a planar antenna array with $32$ antenna elements configured as $16$ patches, each consisting of two antennas with $\pm45^\circ$ polarizations, $7$ dBi gain and $90^\circ$ beamwidth on both horizontal and vertical planes. A crown-shaped antenna array is used in the UE, which has $50$ antenna elements allocated on $25$ patches with the structure similar with those on the eNB antenna array. The horizontal and vertical spacings between the nearest neighboring antennas are $0.5$ wavelength. The down-tilting angle of the eNB antenna array is approximately $7^\circ$.

The measurement was conducted emulating the downlink scenarios, i.e.\ the eNB and the UE were considered as the transmitter (Tx) and the receiver (Rx) respectively. The high-resolution channel parameter estimation algorithm Space-Alternating Generalized Expectation-maximization (SAGE) algorithm \cite{Yin-Fleury-2003-02} was applied to estimation of $14$ parameters for individual paths, i.e.\ the delay, Doppler frequency, direction (azimuth and elevation) of departure (DoD), direction of arrival (DoA) and polarization matrix.

\begin{figure}
\centering
{\small (a) Measurement routes in Area A (Picture: $\copyright$ Baidu Maps)}
\includegraphics[width=7.5cm]{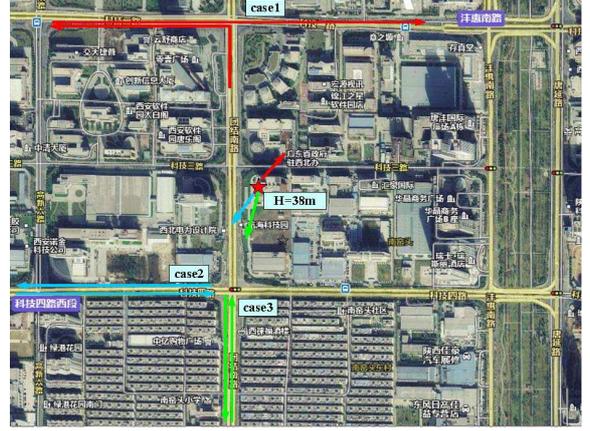}\vspace*{0.3ex}

{\small (b) A photograph taken in Area A}
\includegraphics[width=7.5cm]{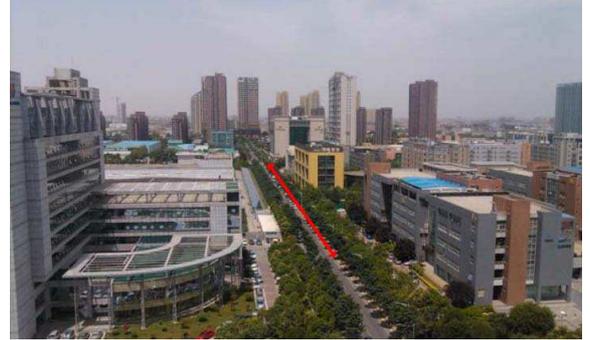}
\caption{Measurement routes and a photograph taken in Area A}\label{figure3}
\end{figure}

\begin{figure}
\centering
{\small (a) Measurement routes in Area B (Picture: $\copyright$ Baidu Maps)}
\includegraphics[width=7.5cm]{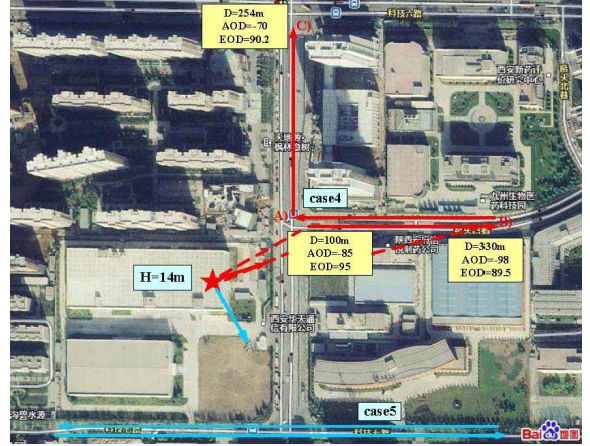}\vspace*{0.3ex}

{\small (b) A photograph taken in Area B}
\includegraphics[width=7.5cm]{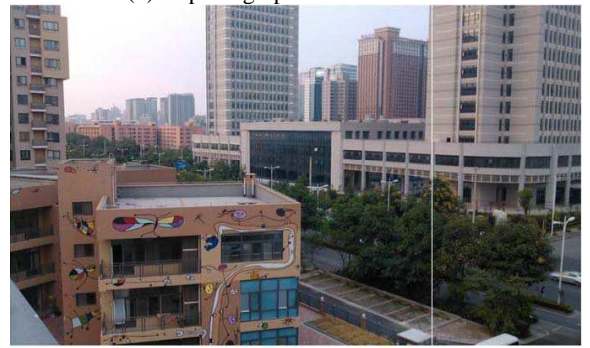}
\caption{Measurement routes and a photograph taken in Area B}\label{figure4}
\end{figure}

The measurements were conducted in the ``High-tech'' district of Xi'an, China. The UE was moving along two routes in the area A and area B as indicated in Figure \ref{figure3} and \ref{figure4} respectively. The photographs taken in the eNB side for the area A and B are shown in Figure \ref{figure3} and \ref{figure4}. It can be observed that the area A and B are close to the definition of UMA and UMI scenarios  specified in the WINNER modeling \cite{WINNER} receptively. The heights of the eNB antenna in area A and area B were $40$ m and $14$ m above the ground respectively. Notice that considering the limited effective radiation of the antenna array in the eNB, the estimation range of azimuth of departure (AoD) and elevation of departure (EoD) are confined to be $[-180^\circ, 0^\circ]$ and $[0^\circ , 180^\circ]$, respectively. Furthermore, the AoD and EoD of the direction normal to the array plane are $-90^\circ$ and $90^\circ$ respectively.

%In order to evaluate the accuracy of parameter estimation, the AoD and EoD of the LoS path at both the start and the end of the measurement route in the ``case 4'' for the area B are  calculated geometrically based on the map.

\section{Verification of the parameter estimation}
\begin{figure}
\centering
\psfrag{a}[c][t][0.9]{(a) Estimated AoD of the LoS path}
\psfrag{b}[c][t][0.9]{(b) Estimated EoD of the LoS path}
\psfrag{Time Snapshot}[c][t][0.9]{Snapshot index}
\psfrag{AOD}[c][t][0.9]{AoD $[^\circ]$}
\psfrag{EOD}[c][t][0.9]{EoD $[^\circ]$}

\hspace*{-4ex}\includegraphics[width=10cm]{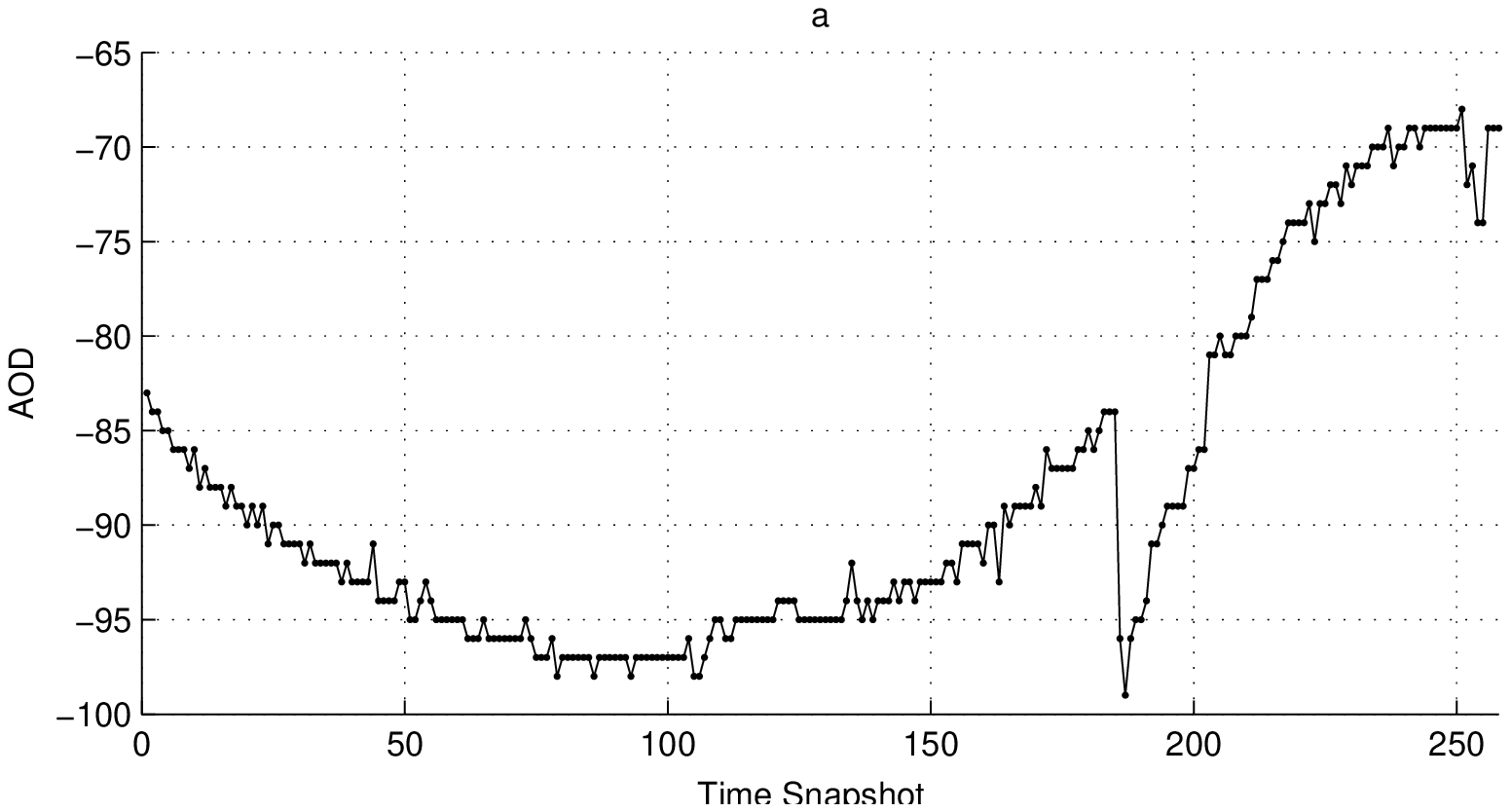}\vspace*{2ex}

\hspace*{-4ex}\includegraphics[width=10cm]{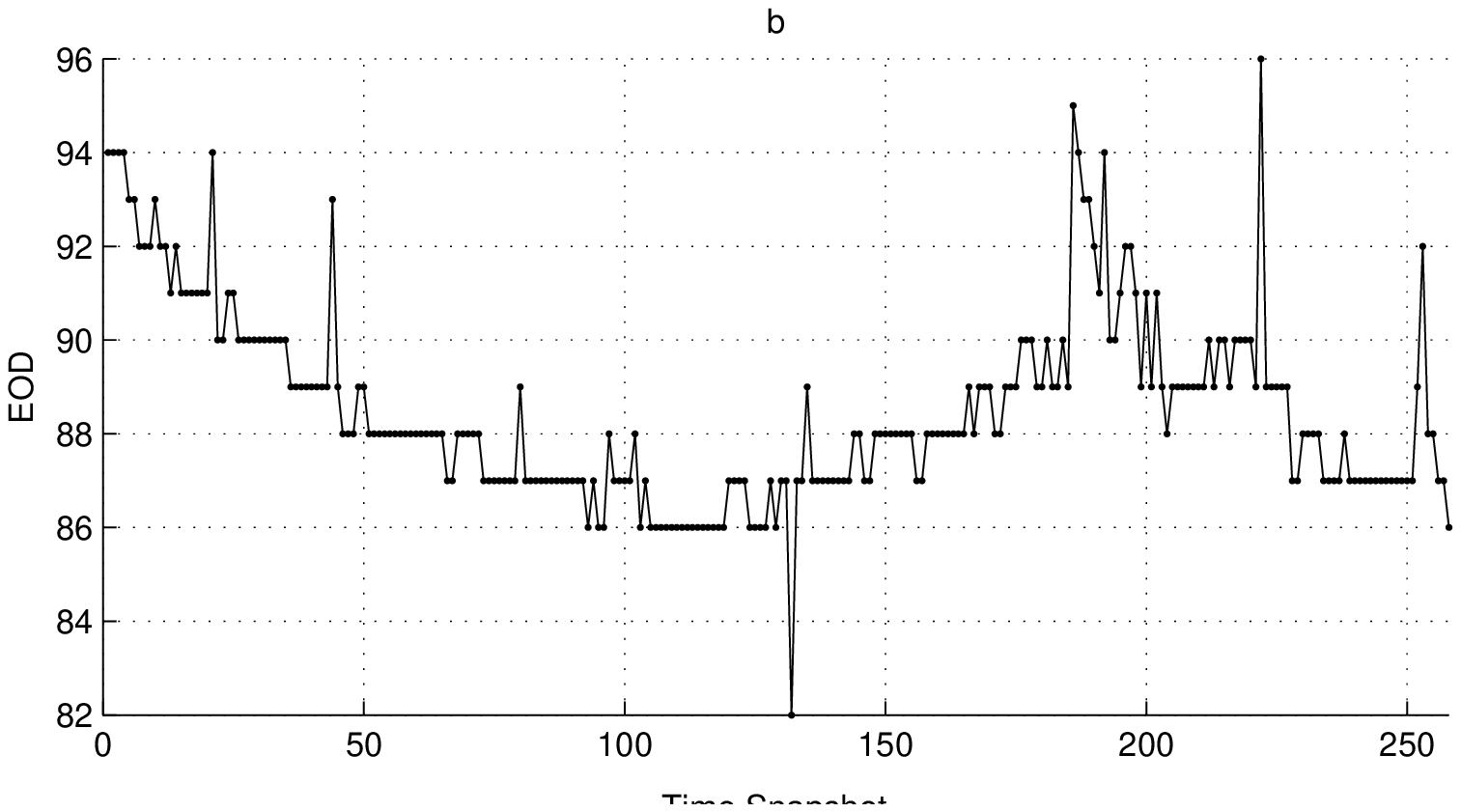}
\caption{Estimated AoD and EoD for LOS path in case 4}\label{figure5}
\end{figure}

To ensure that the path parameters are correctly estimated, the angular estimates obtained in the LoS scenario in the case 4 in the area B are compared with their counterparts calculated geometrically based on the map information and the antenna heights. It can be observed from Figure \ref{figure4} that in the measurement case 4, the UE started from the location A, moving to the location B, then back to A, and finally arrived at the location C. The theoretical AoD and EoD for the LoS path at the start and the end point of this measurement route are illustrated in Figure \ref{figure4}.
Figure \ref{figure5} (a) depicts the AoD estimates for the LoS path, which is selected to be the strongest path estimated by using the SAGE algorithm, versus the measurement snapshots. It can be seen that the AoD estimate increases gradually from $-85^\circ$ to $-98^\circ$, and decreases to $-85^\circ$ when the UE arrived at the location A, and then increase again up to $-70^\circ$ when the UE arrived at the location C. This variation is consistent with the AoD's trajectory calculated theoretically based on the UE displacement. It can also be observed from Figure \ref{figure5} (a) that there is a sudden change of the AoD at the $180$th measurement snapshot. This is due to the fact that the orientation of the Tx antenna array was adjusted manually at that moment in order to make sure that the UE can be covered by the array's main lobe. The array was tuned back to its original position finally. Figure \ref{figure5} (b) depicts the estimated EoDs of the LoS path versus the snapshots, which also fits well with the theoretically calculated EoDs. In addition, it is worth mentioning that the estimated trajectories of AoDs and EoDs are not smooth due to the limited intrinsic resolution of the measurement equipment in both delay and angular domains.

\begin{figure}
\centering
\psfrag{ EOD distribution at Macro scenario}[c][t][0.9]{EoD $[^\circ]$}
\psfrag{micro scenario  EOD distribution}[c][t][0.9]{EoD $[^\circ]$}
\psfrag{pdf}[c][t][0.9]{Probability density function}
\psfrag{macro}[c][c][0.9]{(a) Urban macro LoS scenario}
\psfrag{micro}[c][c][0.9]{(b) Urban micro LoS scenario}
\psfrag{Empirical data}[l][c][0.8]{\hspace*{-6.5ex}Empirical data}
\psfrag{Laplacian pdf}[l][c][0.8]{\hspace*{-6ex}Laplacian pdf}
\includegraphics[width=8cm]{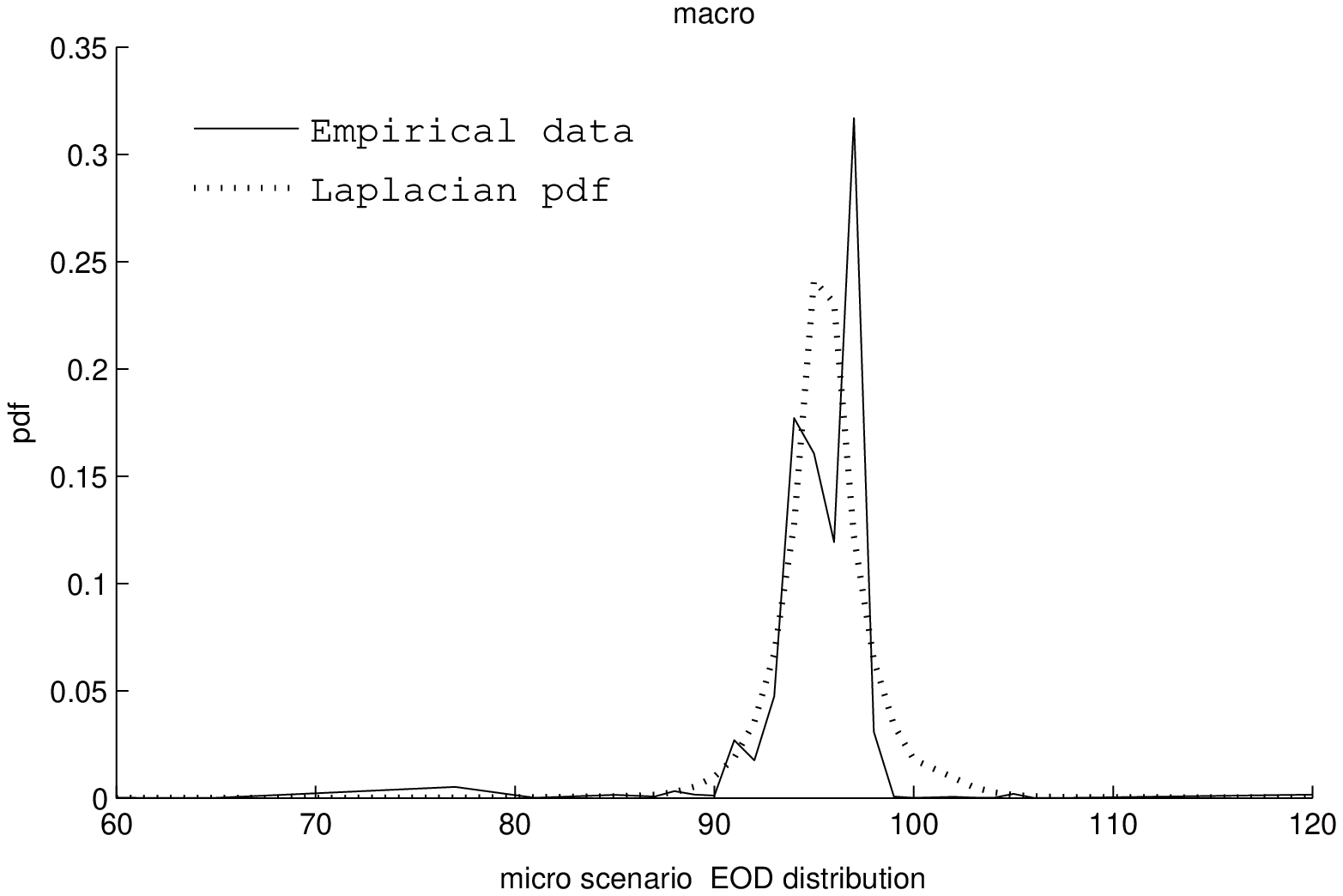}
\includegraphics[width=8cm]{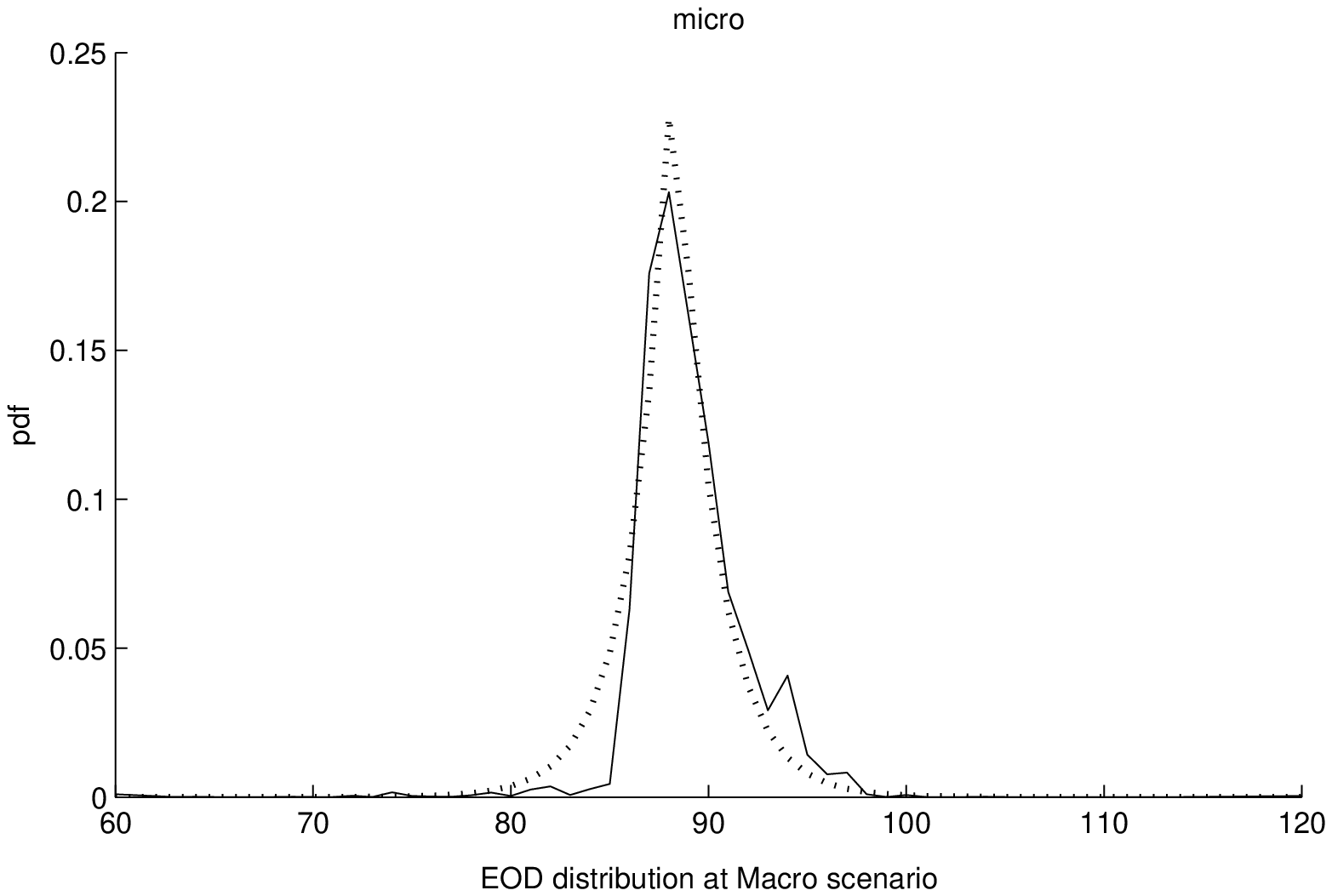}
\caption{EoD distribution in UMA \& UMI LOS scenarios}\label{figure6}
\end{figure}
\section{Measurement results}\label{sect:resultsanalysis}
\subsection{Measurement-based 3D channel parameters}
 The measurements were conducted with the objective of extending the ITU channel model \cite{RP-121412} to include the elevation dimension in order to generate the 3D channel model for system level simulations. Table \ref{t1} reports the statistical channel parameters extracted for both the elevation and azimuth domains at the eNB and UE sides, as well as the delay domains. In the table, DS, ASD and ASA represent the delay spread, azimuth spread for departure and for arrival, respectively. These parameters can be used to update the corresponding entries of the table A1-7 in \cite{M2135} for ITU channel model standards for the environments considered in the measurements.
\begin{table}
\centering \caption{Channel model parameters extracted}
\begin{tabularx}{8.5cm}{|l|X|X|X|X|X|}\hline
\multicolumn{2}{|c|}{\multirow{2}{*{Scenarios}}} & \multicolumn{2}{|c|}{UMI} & \multicolumn{2}{|c|}{UMA} \\ \cline{3-6}
\multicolumn{2}{|c|}{} & LoS & NLoS & LoS & NLoS \\
\hline
\multirow{2}{2cm}{Delay spread (DS) $\log_{10}([s])$} & $\mu$ & $-7.08$ & $-7.6$ & $-6.66$ & -6.51 \\
& $\sigma$ & 0.34	& 0.8	& 0.56	& 0.2313\\
\hline
\multirow{2}{2cm}{ASD $\log_{10}([^\circ])$} & $\mu$ & 0.84 &	1.1&	0.61	 &	1.05 \\
& $\sigma$ & 0.19 &	0.3&	0.32	&0.1\\
\hline
\multirow{2}{2cm}{ASA $\log_{10}([^\circ])$} & $\mu$ & 1.64	&1.5	&1.68	 & 1.87 \\
& $\sigma$ & 0.17&	0.13&	0.18&	0.11\\
\hline
\multirow{2}{2cm}{ESD $\log_{10}([^\circ])$} & $\mu$ & 0.27 &	0.6	& 0.11	& 0.8346\\
& $\sigma$ & 0.29 &	0.2&	0.2&	0.0452 \\
\hline
\multirow{2}{2cm}{ESA $\log_{10}([^\circ])$} & $\mu$ &1.06&	0.88&	1.49&	 1.26\\
& $\sigma$ &0.18&	0.16	&0.43	&0.16 \\
\hline
\multicolumn{2}{|l|}{Mean EoD  $([^\circ])$} & -2	&-2	&2	&-2
\\\hline
\multicolumn{2}{|l|}{Mean EoA  $([^\circ])$}  & 2	&2	&6	&10
\\
\hline
\multirow{2}{2cm}{$K$-factor [dB] } & $\mu$ &9	& N/A 	& 9	& N/A \\
& $\sigma$ &5	& N/A & 	3.5	& N/A  \\
\hline
\multicolumn{2}{|c|}{EoA, EoD distribution} &  \multicolumn{4}{|c|}{Laplacian}\\\hline
\multicolumn{2}{|c|}{Cluster ESD} & 1.45&	3	&1.16	&2.332 \\\hline\multicolumn{2}{|c|}{Cluster ESA  } &6.68&	7	&15.5&	7\\\hline
\multicolumn{2}{|c|}{Per cluster shadowing std  } &3 dB&	3 dB&3 dB	&3 dB\\\hline
\end{tabularx}\label{t1}
\begin{itemize}
\item[]\hspace*{-3ex}ASD: Azimuth Spread of Departure
\item[]\hspace*{-3ex}ASA: Azimuth Spread of Arrival
\item[]\hspace*{-3ex}ESD: Elevation Spread of Departure
\item[]\hspace*{-3ex}ESA: Elevation Spread of Arrival
\item[]\hspace*{-3ex}EoD: Elevation of Departure
\item[]\hspace*{-3ex}EoA: Elevation of Arrival
\end{itemize}
\end{table}

The 3D MIMO performance evaluation also requires the power elevation spectrum and elevation spread at the eNB. Table \ref{t2} reports the mean azimuth angular spread (MAAS) and mean elevation angular spread (MEAS) at the eNB. Figure \ref{figure6} depicts the EoA distributions for the UMA and UMI LoS scenarios. It can be observed from Figure \ref{figure6} that the Laplacian distributions can be used to fit the empirical distribution derived from measurements.

\begin{table}
\centering \caption{Mean azimuth angular spread (MAAS) and mean elevation angular spread (MEAS) at the eNB}
\begin{tabular}{|c|c|c|}
\hline & MAAS $[^\circ]$ &	MEAS $[^\circ]$ \\\hline
UMA LoS	& $4.6$	& $1.77$ \\\hline
UMA NLoS&	$11.2$ &	$6.8$\\\hline
UMI LoS	& $11.54$	& $2.97$ \\\hline
UMI NLoS&	$13.3$	& $6.28$ \\
 \hline
\end{tabular}\label{t2}\vspace*{-3ex}
\end{table}

Table \ref{t3} reports the cross-correlation coefficients between different large-scale parameters (LSPs) based on the measurement results. Here, the cross-correlation coefficient of two random variables, say $a$ and $b$, are calculated as
\begin{align}
C(a,b)=\frac{\sum\limits_{i=1}^N(a_i - \bar{a})(b_i - \bar{b})^*}{\sqrt{\biggl(\sum\limits_{i=1}^N|a_i - \bar{a}|^2\biggr)\biggl(\sum\limits_{i=1}^N|b_i - \bar{b}|^2\biggr)}},\nonumber
\end{align}
where $\bar{a}$ and $\bar{b}$ denote the mean of $a$ and $b$ respectively. The 3D channel realizations can be generated using the parameters given in Tables \ref{t1} to \ref{t3} following the ITU channel generation procedure.
\section{Generating channel coefficient for LOS path}\label{sect:loscoffecients}
In the case where the elevation domain is taken into account, the contribution of the $n$th cluster in the narrowband coefficients of the channel between the $s$th Tx antenna and the $u$th Rx antenna can be calculated as \cite{WINNER}
\begin{align}
\bsH_{u,s,n}(t)=\sqrt{\frac{1}{K_R+1}}\bsH_{u,s,n}'(t)+\delta(n-1)\bsH^{\mathrm{LoS}}_{u,s,n}(t),\label{eq:combining}
\end{align}
where $\bsH_{u,s,n}'(t)$ represents the NLoS component and can be written as (\ref{eq:winner}), the same as in \cite[EQ.(4.20)]{WINNER} in the case where elevation domains are considered. Due to the limitation of space, the readers of interest are referred to  \cite{WINNER} for detailed explanation of the notations arising in (\ref{eq:winner}).
\newcounter{mytempeqncnt}
 \begin{figure*}[!t]
 % ensure that we have normalsize text
 \normalsize
 % Store the current equation number.
 \setcounter{mytempeqncnt}{\value{equation}}
 % Set the equation number to one less than the one
 % desired for the first equation here.
 % The value here will have to changed if equations
 % are added or removed prior to the place these
 % equations are referenced in the main text.
 %\setcounter{equation}{5}
 \begin{align}
\hspace*{-4ex}\bsH_{u,s,n}'(t) &=\!\! \sqrt{P_n}\sum\limits_{m=1}^M\!\left[\begin{matrix}
F_{\mathrm{rx},u,V}(\bar{\Psi}_{n,m}) \\ F_{\mathrm{rx},u,H}(\bar{\Psi}_{n,m}) \end{matrix}\right]^{\mathrm{T}}\!\left[\begin{matrix}\!\!\!
\exp(j\phi_{n,m}^{vv}) &\!\!\!\!\!\!\! \sqrt{\kappa_{n,m}^{-1}}\exp(j\phi_{n,m}^{vh}) \\ \!\! \sqrt{\kappa_{n,m}^{-1}}\exp(j\phi_{n,m}^{hv}) &\!\!\!\!\!\!\!\exp(j\phi_{n,m}^{hh}) \end{matrix}\right]\!
\left[\begin{matrix}
F_{\mathrm{tx},u,V}(\bar{\Phi}_{n,m}) \\ F_{\mathrm{tx},u,H}(\bar{\Phi}_{n,m})\end{matrix}\right]\exp\biggl(j2\pi \bigl(\frac{\bar{\bsr}_s\cdot\bar{\Phi}_{n,m}}{\lambda_0}\!+ \! \frac{\bar{\bsr}_u\cdot\bar{\Psi}_{n,m}}{\lambda_0}\!+ \! \frac{\bar{v}\bar{\Psi}_{n,m}}{\lambda_0}t\bigr)\biggr)\!\!\label{eq:winner}\\
% \end{align}
%  \end{figure*}
%%\newcounter{mytempeqncnt}
% \begin{figure*}[!t]
% % ensure that we have normalsize text
% \normalsize
% % Store the current equation number.
% \setcounter{mytempeqncnt}{\value{equation}}
% % Set the equation number to one less than the one
% % desired for the first equation here.
% % The value here will have to changed if equations
% % are added or removed prior to the place these
% % equations are referenced in the main text.
% %\setcounter{equation}{5}
% \begin{align}
\bsH_{u,s,n}^{LoS}(t) &= \frac{K_R}{K_R+1}\left[\begin{matrix}
F_{\mathrm{rx},u,V}(\bar{\Psi}_{\mathrm{LoS}}) \\ F_{\mathrm{rx},u,H}(\bar{\Psi}_{\mathrm{LoS}}) \end{matrix}\right]^{\mathrm{T}}\left[\begin{matrix}
\exp(j\phi_{\mathrm{LoS}}) & 0 \\ 0 & \exp(j\phi_{\mathrm{LoS}}) \end{matrix}\right]
\left[\begin{matrix}
F_{\mathrm{tx},u,V}(\bar{\Phi}_{\mathrm{LoS}}) \\ F_{\mathrm{tx},u,H}(\bar{\Phi}_{\mathrm{LoS}})\end{matrix}\right]
\exp\biggl(j2\pi\bigl( \frac{\bar{\bsr}_s\cdot\bar{\Phi}_{\mathrm{LoS}}}{\lambda_0}+\frac{\bar{\bsr}_u\cdot\bar{\Psi}_{\mathrm{LoS}}}{\lambda_0}+
\frac{\bar{v}\bar{\Psi}_{\mathrm{LoS}}}{\lambda_0}t\bigr)\biggr)\label{eq:changedLoS}
 \end{align}
  \end{figure*}

\begin{table}
\centering \caption{Cross-correlation of LSPs (DS,ASD,ASA, ESD and ESA ,SF,K)}
\begin{tabularx}{8.3cm}{|c|c|X|X|X|X|}\hline
\multicolumn{2}{|c|}{\multirow{2}{*{Scenarios}}} & \multicolumn{2}{|c|}{UMI} & \multicolumn{2}{|c|}{UMA} \\ \cline{3-6}
\multicolumn{2}{|c|}{} & LoS & NLoS & LoS & NLoS \\
\hline
\multirow{21}{1.3cm}{Cross-Correlations} & ASD vs DS & $0.42$ 	& $0$& 	$0.21$&  	 $0.4$ \\
 & ASA vs DS &	$0.11$ &	$0.4$	&$0.25$ &	$0.6$ \\
 &ASA vs SF &	$-0.4$	&$-0.4$	&$-0.5$&	$0$\\
 & ASD vs SF  &	$-0.5$	 &$0$	&$-0.5$	 &$-0.6$\\
 & DS vs SF 	 &$-0.4$	 &$-0.7$	 &$-0.4$ &	$-0.4$\\
  &ASD vs ASA  &	 $0.11$  &	$0$	 &$0$ & 	$0.4$\\
 &ASD vs $K$   &	$-0.2$	 &N/A  &	$0$ &	N/A\\
 &ASA vs $K$ &  	$-0.3$	  &N/A  &	$-0.2$	 &N/A\\
 &DS vs  $K$ & 	$-0.7$	 &N/A  &	$-0.4$ &	N/A\\
 &SF vs  $K$ & $0.5$ &	N/A  &	$0$	 &N/A\\
 &ESD vs DS  &	$-0.07$	 &$-0.5$	 &$-0.3$ &	$-0.3$\\
 &ESA vs DS  &	$0.12$ &	$0$ &	$0.13$	&$0$\\
 &ESD vs ASD  &	$0$	 &$0.5$ &	$0$	 &$0.3$\\
 &ESA vs ASD  &	$0$ 	 &$0.5$	 &$0.44$ &$-0.2$\\
 &ESA vs ASA  &	$0.37$  &	$0$  &$0.3$	 & $0$\\
 &ESD vs ASA  &	$-0.26$ &	$0$ &	$-0.2$ &	$0$\\
 &ESD vs SF  &	$0$	 &$0$	 &$0$	&$0$\\
 &ESA vs SF  &	$0$ &	$0$	 &$-0.8$ &	$-0.5$\\
 &ESD vs $K$  & 	$0$	 &N/A  &	$0$	 &N/A\\
 &ESA vs $K$  & 	$0$	 &N/A 	 & $0$ &	N/A\\
 &ESD vs ESA  &	$-0.2$ 	 &$0$	& $-0.11$  &	$0$\\
\hline
\multirow{7}{1.3cm}{Correlation distance [m]} &DS& 	7	&10&	30&	40\\
&ASD &	8&	10	&18	&50\\
&ASA &	8	&9	&15&	50\\
&ESD &	8&	9.5&	16.5&	50\\
&ESA &	8	&9.5	&16.5	&50\\
&SF 	&10&	13&	37	&50\\
  &K	&15&	N/A 	&12&	N/A\\
\hline
\end{tabularx}\label{t3}
\begin{itemize}
\item[]SF: Shadow Fading
\item[]DS: Delay Spread
\item[]
\end{itemize}
\end{table}

%
%\begin{equarray*}
%\bsH_{u,s,n}'(t) = \sqrt{P_n}\sum\limits_{m=1}^M\left[\begin{matrix}
%F_{\mathrm{rx},u,V}(\bar{\Psi}_{n,m}) \\ F_{\mathrm{rx},u,H}(\bar{\Psi}_{n,m}) \end{matrix}\right]^{\mathrm{T}}\left[\begin{matrix}
%\exp(j\Phi_{n,m}^{vv}) & \sqrt{\kappa_{n,m}^{-1}}\exp(j\Phi_{n,m}^{vh}) \\ \sqrt{\kappa_{n,m}^{-1}}\exp(j\Phi_{n,m}^{hv}) & \exp(j\Phi_{n,m}^{hh}) \end{matrix}\right]
%\left[\begin{matrix}
%F_{\mathrm{tx},u,V}(\bar{\Phi}_{n,m}) \\ F_{\mathrm{tx},u,H}(\bar{\Phi}_{n,m})\end{matrix}\right]
%\exp(j2\pi \lambda_0^{-1}\bar{\bsr}_s\cdot\bar{\Phi}_{n,m}) \exp(j2\pi \lambda_0^{-1}\bar{\bsr}_u\cdot\bar{\Psi}_{n,m})\exp\bigl(j2\pi \frac{\bar{v}\bar{\Psi}_{n,m}}{\lambda_0}t\bigr)\label{eq:winner}
%\end{equarray*}

In the case of clear LoS without obstructions, $\bsH_{u,s,n}^{LoS}(t)$ is written as (\ref{eq:changedLoS}), where the diagonal elements of the polarization matrix are co-phased, i.e.\ \begin{align}
\phi_{\mathrm{LoS}}^{vv}=\phi_{\mathrm{LoS}}^{hh}=\phi_{\mathrm{LoS}},\nonumber
\end{align}which is different from the settings in the existing ITU and 3GPP geometry-based channel models that $\phi_{\mathrm{LoS}}^{vv}$ and $\phi_{\mathrm{LoS}}^{hh}$ are generated independently. This revision is important since in the case where the LoS path is of free-space propagation, the polarization status of the transmitted EM wave shall be unaffected, which occurs if and only if the diagonal elements of the LoS path polarization matrix do not introduce extra phase difference to the vertical and horizontal polarized components being transmitted. The conventional setting of independent $\Phi_{\mathrm{LoS}}^{vv}$ and $\Phi_{\mathrm{LoS}}^{hh}$ results in a consequence that the polarization status of received LoS component is different from that of the transmitted signal, which is inconsistent with the common experience in reality.

\begin{figure}
\centering
\includegraphics[width=6cm]{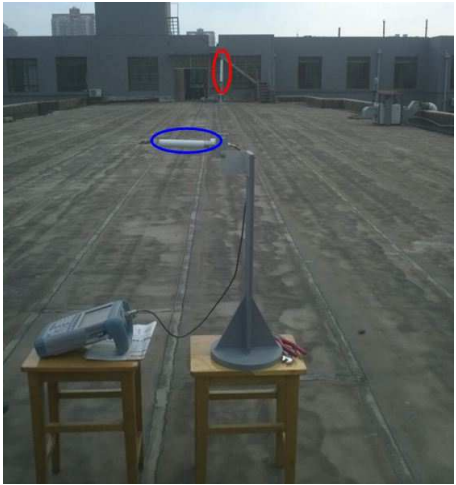}
\caption{Measurement in a LoS environment. The antennas marked by the red and blue ellipses are for the Tx and Rx respectively.}\label{figure7}\vspace*{-2ex}
\end{figure}

To justify the amendment suggested, a measurement campaign was conducted to check the variability of polarization status for the signals received from the LoS path. Figure \ref{figure7} depicts the premises of the measurement in the open rooftop environment with clear LoS between the Tx and Rx separated by $40$ meters. The Tx was equipped with a patch consisting of two antennas transmitting $\pm 45^\circ$ polarized waves respectively. By feeding different phases into the input signals at radio frequency of the two Tx antennas, the transmitted wave may have desirable polarizations, such as the linear, elliptical or circular polarizations with either left-hand or right-hand circulations. For example, the vertical linearly-polarized waves and circularly-polarized waves can be generated by tuning the phase difference to $0^\circ$ and $90^\circ$, respectively. Furthermore, during the measurements, a phasor-monitoring device was applied that makes sure the correct phase difference be maintained between the two branches of transmitted signals. The Rx was equipped with a dipole antenna, which has tunable orientation that allows receiving EM waves with any desirable linear polarizations.

\begin{figure}\vspace*{-3ex}
\centering
\psfrag{circular polarized wave}[l][c][0.9]{\hspace*{-12ex}Circularly-polarized wave being transmitted}
\psfrag{linear polarized wave}[l][c][0.9]{\hspace*{-11ex}Linearly-polarized wave being transmitted}
\psfrag{Receiving Polarization Angle}[c][c][0.9]{Polarization Slant Angle at Rx $[^\circ]$}
\psfrag{Received Power}[c][c][0.9]{Power of the received signal [dB]}

\hspace*{-3ex}\includegraphics[width=10cm]{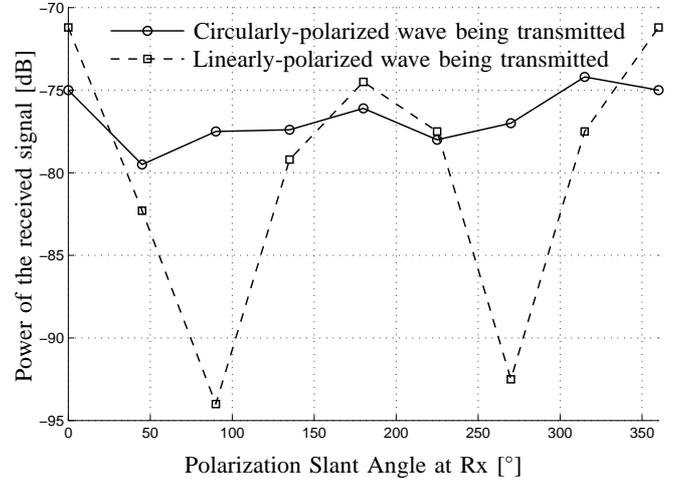}
%\rput[tl](-6cm,3.7cm){\small (a) the Tx}
%\rput[tl](-2.5cm,3.7cm){\small  (b) the Rx}
\caption{Power of the received signal when the Rx polarization varies from $0^\circ$ to $360^\circ$ polarizations}\label{figure8}\vspace*{-2ex}
\end{figure}

Figure \ref{figure8} reports the received power at the output of the Rx with linear polarizations from $0^\circ$ to $360^\circ$. Notice that the polarization slant angle of $0^\circ$ is referred to the vertical polarization direction in the measurement. It can be observed from Figure \ref{figure8} that when the circularly-polarized wave was transmitted, the received power remained stable regardless of the polarizations of Rx antenna. When the vertical linearly-polarized wave was transmitted, i.e.\ the polarization was along $0^\circ$, the received power achieved the maxima when the Rx polarization matched the Tx polarization, i.e.\ being $90^\circ$ or $270^\circ$ polarized, and reduced to the minima when the Rx and Tx polarizations were orthogonal. It is evident that the polarization of the transmitted waves does not change in the LoS scenario considered in the measurement, implying that the diagonal elements of the LoS polarization matrix do introduce the same phase variation to both the vertical and horizontal polarized waves.

%\begin{equation*}
%\bsH_{u,s,n}'(t) = \frac{K_R}{K_R+1}\left[\begin{matrix}
%F_{\mathrm{rx},u,V}(\bar{\Psi}_{n,m}) \\ F_{\mathrm{rx},u,H}(\bar{\Psi}_{n,m}) \end{matrix}\right]^{\mathrm{T}}\left[\begin{matrix}
%\exp(j\Phi_{\mathrm{LoS}}) & 0 \\ 0 & \exp(j\Phi_{\mathrm{LoS}}) \end{matrix}\right]
%\left[\begin{matrix}
%F_{\mathrm{tx},u,V}(\bar{\Phi}_{\mathrm{LoS}}) \\ F_{\mathrm{tx},u,H}(\bar{\Phi}_{\mathrm{LoS}})\end{matrix}\right]
%\exp(j2\pi \lambda_0^{-1}\bar{\bsr}_s\cdot\bar{\Phi}_{\mathrm{LoS}}) \exp(j2\pi \lambda_0^{-1}\bar{\bsr}_u\cdot\bar{\Psi}_{\mathrm{LoS}})\exp\bigl(j2\pi \frac{\bar{v}\bar{\Psi}_{\mathrm{LoS}}}{\lambda_0}t\bigr)\label{eq:changedLoS}
%\end{equation*}

\section{Conclusions}\label{Sect:conclusions}
In this contribution, the stochastic channel characteristics extracted based on measurements for urban macro and micro scenarios in Xi'an, China have been reported with emphasis in the elevation of arrival and of departure domains. The elevation angular spreads and the cross-correlation coefficients thereof with other large-scale channel parameters were calculated and compared with the existing 3-dimensional (3D) ITU IMT-Advanced channel models for urban scenarios. The channels generated by using the updated model parameters yield non-negative definite correlation matrices which is difficult to obtain by using the existing models. In addition, modification were proposed for the conventional settings of the LoS-path polarization matrix defined in the WINNER+/ITU/3GPP spatial channel models. Measurement results justified the reasonability of the changes for real LoS cases.
\bibliography{references}

% Generated by IEEEtran.bst, version: 1.13 (2008/09/30)
\begin{thebibliography}{1}
\providecommand{\url}[1]{#1}
\csname url@samestyle\endcsname
\providecommand{\newblock}{\relax}
\providecommand{\bibinfo}[2]{#2}
\providecommand{\BIBentrySTDinterwordspacing}{\spaceskip=0pt\relax}
\providecommand{\BIBentryALTinterwordstretchfactor}{4}
\providecommand{\BIBentryALTinterwordspacing}{\spaceskip=\fontdimen2\font plus
\BIBentryALTinterwordstretchfactor\fontdimen3\font minus
  \fontdimen4\font\relax}
\providecommand{\BIBforeignlanguage}[2]{{%
\expandafter\ifx\csname l@#1\endcsname\relax
\typeout{** WARNING: IEEEtran.bst: No hyphenation pattern has been}%
\typeout{** loaded for the language `#1'. Using the pattern for}%
\typeout{** the default language instead.}%
\else
\language=\csname l@#1\endcsname
\fi
#2}}
\providecommand{\BIBdecl}{\relax}
\BIBdecl

\bibitem{RP-121412}
\emph{{Study on 3D-channel model for Elevation Beamforming and Massive MIMO
  studies for LTE}}, 3GPP RP-121412 Std.

\bibitem{TR25996}
\emph{Spatial channel model for Multiple Input Multiple Output (MIMO)
  simulations,}, 3GPP TR 25.996 Std.

\bibitem{M2135}
\emph{Guidelines for evaluation of radio interface technologies for
  IMT-Advanced (12/2009)}, ITU-R M.2135-1 Std.

\bibitem{WINNER}
\emph{WINNER II interim channel models}, IST-4-027756 WINNER D1.1.1 Std.

\bibitem{WINNER+}
\emph{D5.3: WINNER+ Final Channel Models}, P. Heino, ed. Std.

\bibitem{Kalliola1}
K.~Kalliola, H.~Laitinen, P.~Vainikainen, M.~Toeltsch, J.~Laurila, and
  E.~Bonek, ``3-d double-directional radio channel characterization for urban
  macrocellular applications,'' \emph{IEEE Trans. On Antennas and
  Propagations}, vol.~51, no.~11, nov. 2003.

\bibitem{Kalliola2}
K.~Kalliola, K.~Sulonen, H.~Laitinen, O.~Kivekas, J.~Krogerus, and
  P.~Vainikainen, ``Angular power distribution and mean effective gain of
  mobile antenna in different propagation environments,'' \emph{IEEE Trans. On
  Vehicular Technology}, vol.~51, no.~5, sep 2002.

\bibitem{Ericsson1}
J.~Medbo, H.~Asplund, J.-E. Berg, and N.~Jalden, ``Directional channel
  characteristics in elevation and azimuth at an urban macrocell base
  station,'' in \emph{Antennas and Propagation (EUCAP), 2012 6th European
  Conference on}, march 2012, pp. 428 --432.

\bibitem{Yin-Fleury-2003-02}
X.~Yin, B.~H. Fleury, P.~Jourdan, and A.~Stucki, ``Polarization estimation of
  individual propagation paths using the {SAGE} algorithm,'' in
  \emph{Proceedings of the IEEE International Symposium on Personal, Indoor and
  Mobile Radio Communications (PIMRC)}, Beijing, China, Sep. 2003.

\end{thebibliography}
\end{document}